\documentclass{article} % For LaTeX2e
\usepackage{iclr2023_conference,times}

\usepackage{amsmath,amsfonts,bm}

\def\eqref#1{equation~\ref{#1}}
\def\1{\bm{1}}

\DeclareMathAlphabet{\mathsfit}{\encodingdefault}{\sfdefault}{m}{sl}
\SetMathAlphabet{\mathsfit}{bold}{\encodingdefault}{\sfdefault}{bx}{n}

\usepackage{caption}
\usepackage{subcaption}
\usepackage{hyperref}
\usepackage{url}
\usepackage{graphicx}

\title{NELoRa-Bench: A Benchmark for Neural-enhanced LoRa Demodulation\thanks{Accepted by International Conference on Learning Representations (ICLR'23) Workshop on Machine Learning for IoT.}}
\author{Jialuo Du$^1$, Yidong Ren$^2$, Mi Zhang$^3$, Yunhao Liu$^1$, Zhichao Cao$^2$\\
$^1$Tsinghua University $^2$Michigan State University $^3$The Ohio State University\\
\texttt{dujl20@mails.tsinghua.edu.cn, renyidon@msu.edu, mizhang.1@osu.edu,}\\
\texttt{yunhao@tsinghua.edu.cn, caozc@msu.edu} 
}

\iclrfinalcopy % Uncomment for camera-ready version, but NOT for submission.
\begin{document}

\maketitle

\begin{abstract}
Low-Power Wide-Area Networks (LPWANs) are an emerging Internet-of-Things (IoT) paradigm marked by low-power and long-distance communication. 
Among them, LoRa is widely deployed for its unique characteristics and open-source technology.
By adopting the Chirp Spread Spectrum (CSS) modulation, LoRa enables low signal-to-noise ratio (SNR) communication. The standard LoRa demodulation method accumulates the chirp power of the whole chirp into an energy peak in the frequency domain. In this way, it can support communication even when SNR is lower than -15 dB.
Beyond that, we proposed NELoRa~\cite{nelora}, a neural-enhanced decoder that exploits multi-dimensional information to achieve significant SNR gain. 
This paper presents the dataset used to train/test NELoRa, which includes 27,329 LoRa symbols with spreading factors from 7 to 10, for further improvement of neural-enhanced LoRa demodulation. The dataset shows that NELoRa can achieve 1.84--2.35 dB SNR gain over the standard LoRa decoder. The dataset and codes can be found at \url{https://github.com/daibiaoxuwu/NeLoRa\_Dataset}.
\end{abstract}

\section{Introduction}
\label{sec-introduction}

Recent years have witnessed the emergence of Low-Power Wide-Area Networks (LPWANs) as a promising mechanism to connect billions of low-cost Internet of Things (IoT) devices for wide-area data collection (e.g., smart-industry, smart-city, smart-agriculture)~\citep{ma2021bond, liu2020pushing}.
Long Range (LoRa)~\citep{LoRaWAN}, SIGFOX~\citep{SIGFOX}, and NB-IoT~\citep{NBIoT} are the three commercialized wireless technologies that facilitate the establishment of LPWANs. 
Among them, LoRa is the only open-source one and works on unlicensed frequency bands. 
By modulating data via Chirp Spread Spectrum (CSS), LoRa allows sensor nodes to send data at low data rates to gateways several or even tens of miles away. 
Theoretically, the CSS mechanism expands each LoRa symbol to a long time period, and the signal power in this time period can be condensed in the frequency domain by the dechirp process during demodulation, constructively adding up into an energy peak, while the noise can only add up destructively, thus raising the energy peak of the signal over the noise floor even in extremely low signal-to-noise ratio (SNR).

Unfortunately, recent studies~\citep{eletreby_empowering_2017,dongare_charm_2018,gadre_quick_2020,gadre_frequency_nodate,deepLora,Losee,SateLoc,liando2019known,iova2017lora,demetri2019automated,ren2022lorawan} show that the communication range of LoRa is far from the expectation in complex real-world environments (e.g., urban areas, campus). 
The blockage attenuation could severely degrade the SNR of LoRa packets, causing decoding failures even at a sub-kilometer distance. 
Intuitively, if we can improve the LoRa demodulation methods to upgrade its decoding success rate at low SNR, the communication range will be enlarged, increasing the usability of the LPWANs, e.g., extended battery lifetime or reduced number of gateways~\citep{ren2022lorawan,gadre_frequency_nodate,balanuta_cloud-optimized_2020,eletreby_empowering_2017}.

The standard LoRa demodulation process of the dechirp has not reached the optimal SNR tolerance. The main reason is that the dechirp distinguishes the LoRa signal from noise by only examining the energy in the frequency domain, and might miss valuable information in the time domain. 
Neural networks, however, are suitable for efficiently extracting such multi-dimensional information. We proposed NELoRa~\cite{nelora}, a neural-based LoRa decoding method, achieving significant SNR gain. 

In this paper, we present the dataset used for training and testing NELoRa. It consists of LoRa symbols captured in indoor environments with a USRP N210 Software Defined Radio (SDR). The LoRa packets are preprocessed by locating their headers, removing their Carrier Frequency Offsets (CFO) by the preambles, and slicing each LoRa symbol into a single file labeled with its packet ID and its ground truth. The dataset consists of 4 spreading factors: 7, 8, 9, and 10, to support LoRa decoding experiments on different spreading factors. The LoRa symbols have a bandwidth of 125K. The dataset consists of 27,329 symbols in total. With the dataset, we show that NELoRa can achieve 1.8-2.35~dB SNR gain compared to the dechirp.
The dataset can be found at \url{https://github.com/daibiaoxuwu/NeLoRa\_Dataset}.
This dataset can serve as a benchmark to inspire future research on all kinds of novel demodulation methods on LoRa signals.

\section{Understanding the Problem}
\label{sec-background}

\begin{figure}[!t]
    \centering
    \includegraphics[width=0.7\linewidth]{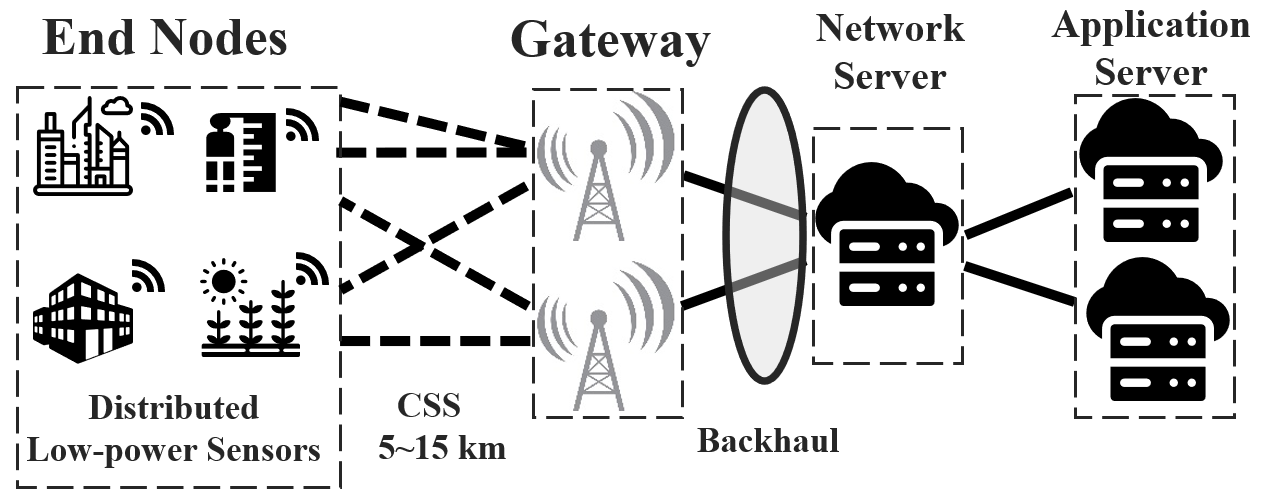}
    \caption{Overview of LoRaWAN architecture.}
    %\vspace{-3mm}
    \label{fig:lora}
    %\vspace{-3mm}
\end{figure}

\begin{figure}
    \centering
    \includegraphics[width=0.7\textwidth]{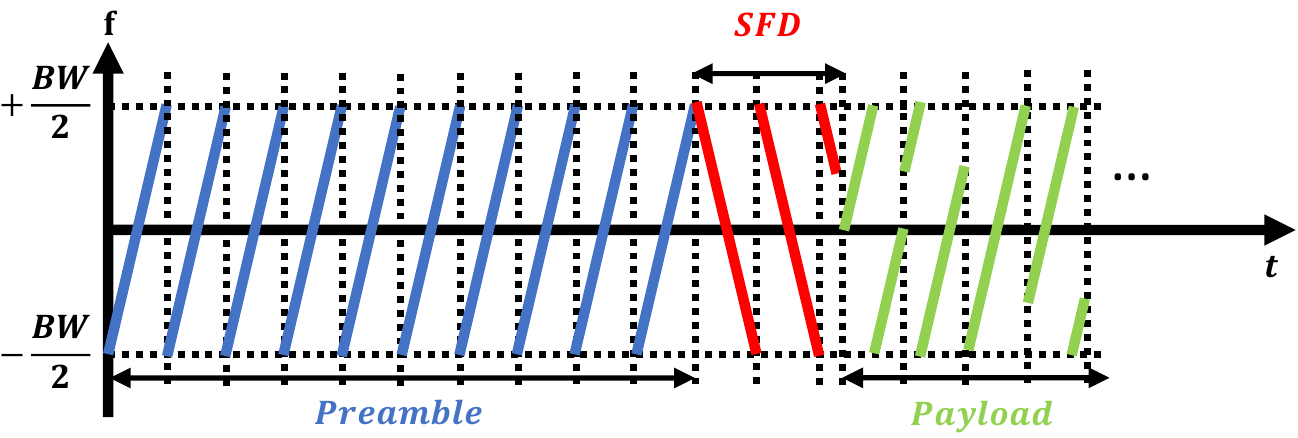}
    \caption{The illustration of a LoRa packet.}
    \label{fig:packet}
\end{figure}

\begin{figure}[!t]
    \centering
    \includegraphics[width=1\linewidth]{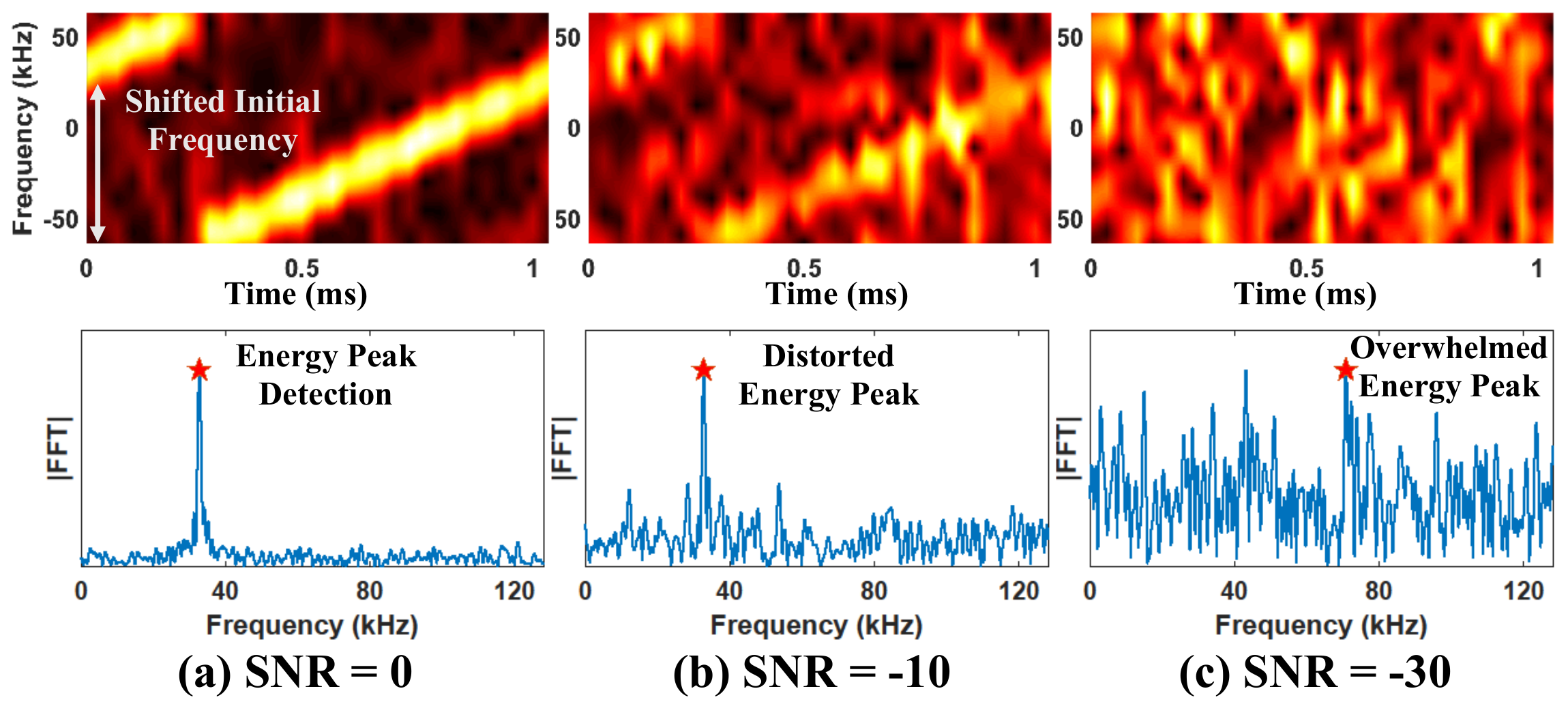}
    %\vspace{-3mm}
    \caption{In dechirp, the energy peak of a chirp symbol's spectrum is distorted or overwhelmed as the SNR decreases.}
    \label{fig:dechirp-processing}
    %\vspace{-3mm}
\end{figure}

As illustrated in Figure~\ref{fig:lora}, a LoRaWAN consists of end nodes, gateways, a network server, and an application server. The collected sensory data (e.g., temperature, humidity) transmitted from the distributed end nodes is relayed by several gateways to the network server. A standard LoRa packet consists of 3 parts: preamble, start frame delimiter (SFD), and payload, as illustrated in Fig.~\ref{fig:packet}. LoRa uses CSS modulation~\cite{berni1973utility}. A base up-chirp is a signal whose frequency increases linearly over time from $-\frac{BW}{2}$ to $\frac{BW}{2}$, and the preamble consists of eight of them. The SFD consists of two and a quarter of base down-chirps, which is the conjugate of up-chirp, whose frequency decreases linearly over time from $\frac{BW}{2}$ to $-\frac{BW}{2}$. The preamble is mainly used for packet detection, and the SFD assists the preamble in calibrating frequency offsets during transmission. The payload is the part that contains data. The data bits are then encoded by shifting the initial frequency of a base up-chirp to $f_{s}$, so the frequency increases linearly from $f_{s}$ to $\frac{BW}{2}$, jumps to $-\frac{BW}{2}$, and then increases linearly from $-\frac{BW}{2}$ to $f_{s}$. During LoRa demodulation, after packet detection and CFO correction by the preamble and SFD~\cite{nelora,tong_colora_2020}, we can slice out each symbol and detect its initial frequency to decode which data it contains. LoRa further defines a configuration parameter called $SF$, and divides the bandwidth $-\frac{BW}{2}$ to $\frac{BW}{2}$ into $2^{SF}$ frequencies, and the initial frequency of each data bit falls in one of these frequencies. Thus, each symbol can encode $SF$ bits, and there are a total of $2^{SF}$ different LoRa symbols in total. The decoding of LoRa symbols is actually a classification problem where we classify the received symbol as one of the $2^{SF}$ symbols.

The standard way for decoding LoRa packets is the dechirp, shown in Fig.~\ref{fig:dechirp-processing}. It first multiplies the chirp symbol with a time-aligned base down-chirp, and performs Fast Fourier Transform (FFT) on the result. The FFT has $2^{SF}$ bins, and the signal energy will be perfectly condensed into one of the bins, resulting in a high signal peak if there is little noise (see Fig.\ref{fig:dechirp-processing}a), and we can determine the signal by the highest FFT peak; Even if there is a certain degree of noise, the noise energy will be scattered randomly into all the bins, thus making the signal peak stand out high above the noise floor (see Fig.\ref{fig:dechirp-processing}b). However, if SNR becomes overwhelmingly low, the energy peak can be distorted or even overwhelmed by the noise energy, and the dechirp method gives a wrong answer (see Fig.\ref{fig:dechirp-processing}c).

The dechirp method is efficient as it condenses the energy into a peak in the frequency domain. However, this method might miss important information in the time domain. We propose NELoRa~\cite{nelora} to formulate the decoding as a classification problem. We utilize neural networks to extract multi-dimensional features from LoRa symbols to achieve SNR gain. We run a neural network on gateways or network servers to enhance the gateways' decoding abilities. By leveraging the gateway's tolerance on power consumption and its extra compute resources, NELoRa adopts deep learning techniques for weak chirp symbol decoding and increases the end nodes' communication range and battery life. Furthermore, LoRaWAN communications often consist of periodic sensory data, which seldom require low transmission delays, so the time overhead induced by the neural networks will not affect the usability of the LoRaWAN system.

\section{Dataset}
\subsection{Dataset Collection}

\begin{figure}[!t]
    \centering
    \includegraphics[width=0.8\linewidth]{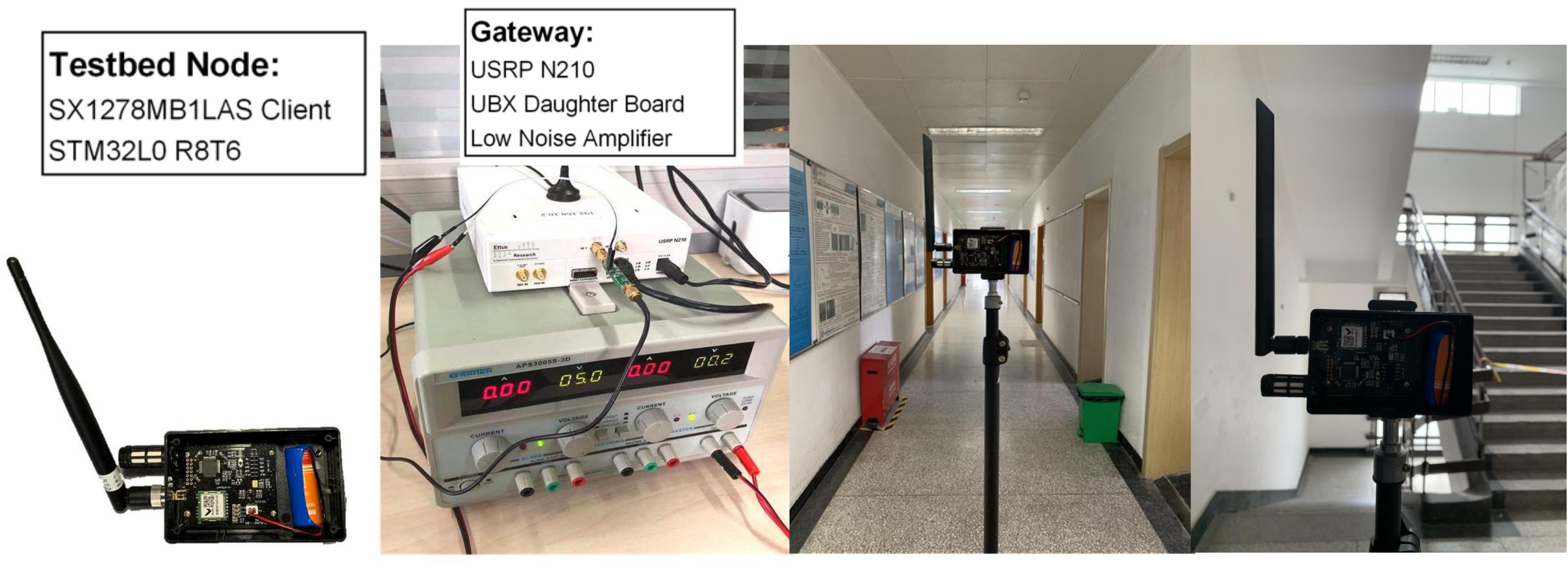}
%    \vspace{-3mm}
    \caption{USRP N210 based gateway and commodity SX1278 client radio based LoRa node.}
    \label{fig:software-hardware}
%    \vspace{-3mm}
\end{figure}

%\vspace{1mm}
\noindent
\textbf{Implementation:} 
%%todo change figure
Figure~\ref{fig:software-hardware} illustrates our data collection system. Specifically, we use the USRP N210 SDR platform for capturing over-the-air LoRa signals, operating on a UBX daughter board at the 470MHz bands and a sampling rate of 1MHz. The captured signal samples are then delivered to a back-end host for pre-processing and demodulation.
On the transmitter side, we use SX1278 client radio based commodity LoRa nodes for transmitting LoRa packets. 

\vspace{1mm}
\noindent
\textbf{Chirp Symbol Dataset:}
We collect LoRa packets at the high SNR ($>30$~dB), including 4 SFs (e..g, 7, 8, 9, 10). Each packet contains around 60 symbols, and we preprocess and slice them into individual symbols. For training and testing, we measure the signal amplitude and add corresponding random-generated Gaussian white noise to render chirp symbols at different SNR, covering -40~dB to 15~dB. 

\textbf{File structure:}
The dataset is contained in 4 folders, one for each SF configuration. Inside them, there are around 100 subfolders, each indicating one packet. Inside each subfolder is around 60 files which contain the I/Q samples for each LoRa symbol, represented as a binary 1-dimensional array of 32-bit float numbers (two consecutive float numbers represent one I/Q sample, with the former and the latter as the real and imaginary part). The ground truth symbol of this file is written in its filename: each datafile's filename is four numbers separated with underscores, indicating 1) the position of the symbol in the packet (starting with 0); 2) The ground truth of this symbol (ranging from $0$ to $2^{SF}-1$); 3) the ID of the packet that contains this symbol (remain the same in each subfolder); 4) the spreading factors (ranging from 7 to 10). The code for the data extracting is presented in the Python file data\_loader.py alongside the dataset.

\vspace{1mm}
\noindent
\subsection{Packet detection}

The default packet detection method utilizes the preamble of a LoRa packet, which consists of multiple continuous base up-chirps. 
If we apply dechirp on the preamble, several continuous energy peaks appear at FFT bin 0 of the multiple base up-chirps' spectrum. 
In practice, a gateway continuously applies dechirp on recorded symbol-length signals (called \emph{window signals}).
If a LoRa preamble appears, a window signal contains a chirp symbol (called \emph{window chirp}) which may not be exactly time-aligned with the base up-chirps in the LoRa preamble. 
Considering the multiple continuous base up-chirps in a LoRa preamble, we will observe several identical window chirps.
If the energy peaks of several window signals appear at the same FFT bin, a LoRa packet is detected.

\subsection{CFO correction}
Due to CFO effects, the frequencies of the LoRa packets may be distorted, hindering proper LoRa decoding. This effect can be mitigated by the SFD portion of the LoRa packets. The key idea is as follows: The LoRa headers contain several base upchirps followed by several base downchirps, and the CFO effects will modify their frequencies in the same way, supposing that the CFO remains constant in the LoRa packet header. Thus, when we demodulate the preamble and SFD by multiplying them with base downchirp and base upchrip, the frequency shifts caused by the CFO on the preamble and SFD will be opposite to each other, where we can measure and mitigate this CFO effect. Thus, with the base down-chirps in the packet's SFD, we remove CFO~\citep{tong_combating_2020} to generate high-quality chirp symbols in the packet. The detailed preprocessing methods and codes are the same as NELoRa~\citep{nelora} and are presented as MATLAB code alongside the dataset. After removing CFO in the whole packet, we can remove the header and slice the payload into individual symbols, which we present in our dataset.

\section{Testing Algorithms}

After constructing our dataset, we implement two methods for LoRa symbol demodulation: 1) the baseline method used in standard LoRa protocol based on dechirp, and 2) our neural-enhanced method NELoRa\cite{nelora,li2022nelora}. The demodulation of LoRa symbols is actually a classification problem. Each LoRa symbol is an up-chirp with varying initial frequencies, which contains information about this symbol. The LoRa protocol divides the bandwidth into $2^{SF}$ frequencies, with $SF$ denoting the spreading factor, and there are a total of $2^{SF}$ LoRa symbols, each taking one of the frequencies as the initial frequency of its upchirp. Consequently, the LoRa decoding problem becomes a classification problem with $2^{SF}$ classes. Consequently, we can use neural-network-based methods to perform this kind of classification.

\textbf{NELoRa Structure:}
\label{sec:nelora_structure}
We implement the neural network model in NELoRa\cite{nelora}, and train and test it using our dataset. We synthetically add Gaussian White Noise on the input signals to produce more training data at different levels of SNR for training and testing. We first apply Short Time Fourier Transform (STFT) on the LoRa symbol, converting the input symbol into a spectrogram. This spectrogram is a 2-dimensional complex-numbered matrix, and we view it as a 2-channel image, separating the real and imaginary parts as two channels. This way, we can apply CNN to this data. We train a denoising CNN that accepts the spectrogram of the synthetic low-SNR signal (added Gaussian White Noise) and uses the corresponding original signal as ground truth, and this CNN will generate a noise-free spectrogram as output. At the same time, we train another CNN network for classification: the input of this network is the output of the denoising CNN, and the output of this network is a vector with a length of $2^{SF}$, indicating the classification result. The two networks are trained simultaneously, and the loss of the two networks, the image MSE loss for the denoising CNN and the classification loss for the classification CNN, are added and backpropagated during training. The detailed description and code are available at NELoRa\cite{nelora} alongside the dataset. We train an individual DNN model for each LoRa spreading factor configuration based on the chirp symbol dataset with the corresponding configuration. 
We further split the dataset into training and test sets. One containing 80\% chirp symbols is used for the DNN model training. The test set includes the rest 20\% of chirp symbols. This test set is also used to test the dechirp method for a fair comparison, which we explain in the following section.

\section{Evaluation}
\subsection{Experiment Settings}
After implementing the two LoRa demodulation methods: dechirp and NELoRa, we now evaluate them on our dataset. We measure the signal amplitude of each symbol and add corresponding degrees of Gaussian White Noise onto the symbols to generate data with different levels of SNR, and evaluate the signal error rate (SER) of the two methods on different levels of SNR.

\begin{figure}
     \centering
     \begin{subfigure}[b]{0.45\textwidth}
         \centering
         \includegraphics[width=\textwidth]{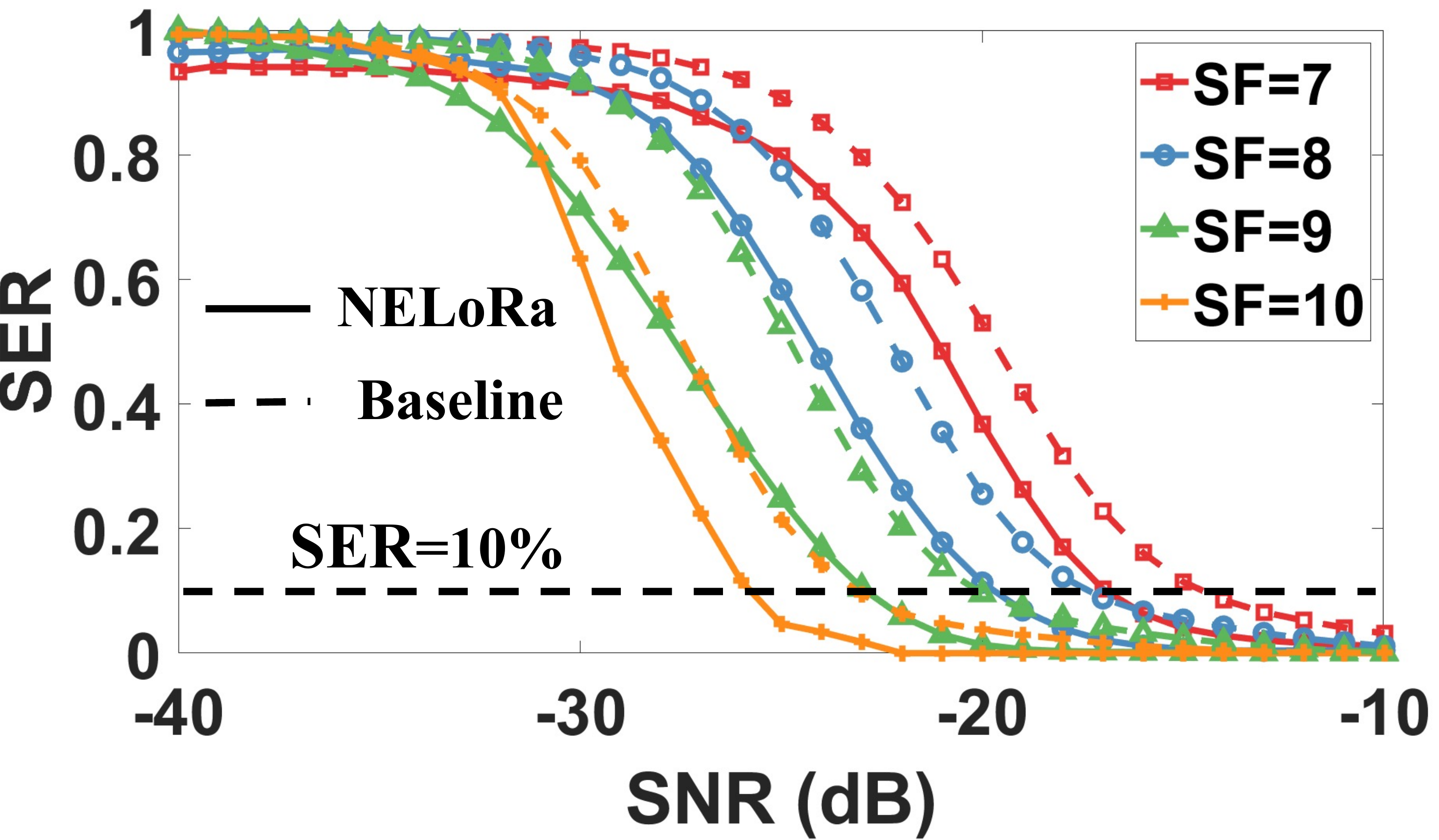}
         \caption{SER-SNR Curves of Different SFs}

     \end{subfigure}
     \hfill
     \begin{subfigure}[b]{0.45\textwidth}
         \centering
         \includegraphics[width=\textwidth]{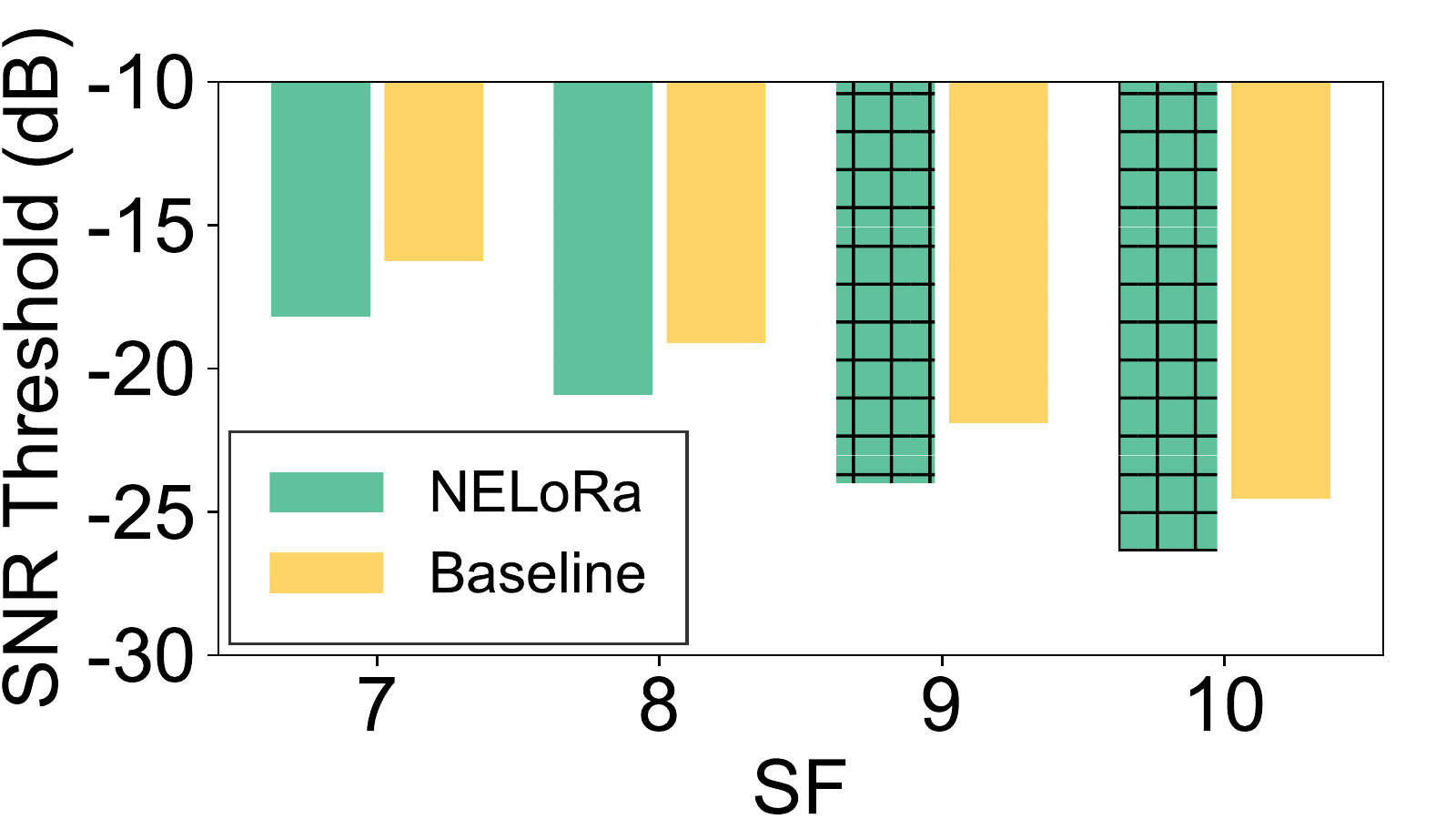}
         \caption{SNR Gains of Different SFs}

     \end{subfigure}
     \hfill
    
        \caption{Overall performance of NELoRa under different LoRa SF configurations. (a) illustrate the SER-SNR curves under different SFs. The solid line and dashed line represent the performance of NELoRa and dechirp, respectively. (b) illustrate the SNR gains under different SFs, with the SNR threshold set at SER=10\%.}
    \label{fig:overall-performance}
\end{figure}

\subsection{Results}
 The results are shown in Figure~\ref{fig:overall-performance}. We can observe that NELoRa (e.g., solid line) has obtained consistently lower SER than dechirp (e.g., dashed line) for SFs from 7 to 10 across all SNR levels. 
For different SFs, the SNR gain is ranging from 1.84~dB (e.g., SF=8) to 2.35~dB (e.g., SF=7)
The results verify the efficiency of our DNN demodulator in ultra-low SNR. And multi-dimensional pattern features are successfully abstracted during the model training process with millions of chirp symbols. 
Our DNN model can be further refined as more diverse chirp symbols are used for training.

\section{Conclusion}
This paper presents a comprehensive dataset of LoRa symbols, covering a spreading factor of 7 to 10 and containing 27,329 symbols. We collect the data on real-life devices and perform thorough preprocessing to detect packets, remove CFO, and slice the payload into individual packets. We further implement two methods: the baseline method dechirp and the neural-based method NELoRa, and evaluate their performances on our dataset. We anticipate that our dataset will support research on new LoRa demodulation methods, especially neural-based ones, and we hope that our dataset can become a benchmark for the evaluation of future LoRa demodulation methods.

\section*{Acknowledgement}
This study is supported in part by NSF Awards CNS-1909177 and NSFC Grant 61972218.

\bibliography{reference}
\bibliographystyle{iclr2023_conference}   

\end{document}